\documentclass[journal]{IEEEtran}
\usepackage{cite}
\usepackage{graphicx}
\usepackage{color}
\usepackage{soul}
\usepackage{ulem}
\usepackage{multicol}
\usepackage{amsmath}
\usepackage{epstopdf}

\begin{document}
\title{Mutual Radiation Impedance of \\ Uncollapsed CMUT Cells with Different Radii}
\author{A. Ozgurluk$^1$, H.K. Oguz$^2$, A. Atalar$^3$, H. Koymen$^3$\\
$^1$ EECS Department, UC Berkeley, CA, USA\\ $^2$ EE Department, Stanford University, CA, USA\\ $^3$ EE Department,
Bilkent University, Ankara, Turkey}

\maketitle
\begin{abstract}
A polynomial approximation is proposed for the mutual acoustic
impedance  between uncollapsed capacitive micromachined ultrasonic
transducer (CMUT) cells with different radii in an infinite rigid
baffle. The resulting approximation is employed in simulating
CMUTs with a circuit model. A very good agreement is obtained with
the corresponding finite element simulation (FEM) result.
\end{abstract}

\section{Introduction}
An array of transducers is usually employed for generating
powerful and focused ultrasonic beams for imaging or therapy
purposes~\cite{yamaner}. Mutual interactions between the cells of the
array are especially important when the mechanical impedance of
cells is relatively low. This is exactly the case for capacitive
micromachined transducers (CMUT) which provide a wide band
operation in liquid immersion because of its low mechanical
impedance~\cite{Park2010}. Mechanics of uncollapsed CMUT cells can
be accurately modelled with flexural disks with clamped
edges~\cite{Ladabaum98KY}. It was shown
that~\cite{koymen12a,oguz12} it is possible to accurately model an
array of CMUT cells using an electrical circuit model coupled to a
impedance matrix composed of self and mutual radiation impedances.
Self and mutual radiation impedances of equal size CMUT cells in
uncollapsed~\cite{senlik10} and collapsed~\cite{ozgurluk} mode can
found in the literature.

Porter~\cite{Porter} was the first to quantify the mutual
impedance between two identical flexural disks for different edge
conditions and obtained an infinite series solution.
Chan~\cite{Chan} extended this work to cover the mutual impedance
of flexural disks with different radii. In both cases, the mutual
impedance turns out to be an oscillatory and slowly decaying
function of the  distance between the two radiators. This implies
that the acoustic coupling between all cells must be taken into
account to accurately predict the performance of an array. This
demanding requirement increases the computation time
substantially. Therefore, a sufficiently accurate but
computationally low-cost approximation is highly desirable to
expedite the simulations of mutual interactions.

An approximation for the radiation impedance of the uncollapsed
CMUT cells of equal size was presented earlier~\cite{oguz12}. This
approximation is extended to that of unequal size cells. It is
verified in Section III performing a finite element method (FEM)
simulation for a particular case.

\section{A Polynomial Approximation}
In Chan's work~\cite{Chan}, the mutual impedance between two
flexural disks of clamped edges with radii $a_1$ and $a_2$ and
center-to-center separation of $d$ is found as\footnote{There is
an additional factor of 9 in (\ref{eq2}), since the reference is
{\it rms} velocity rather than peak velocity~\cite{Chan}.}
\begin{equation}\label{eq2}
\begin{split}
\frac{Z_{12}(ka_1,ka_2,kd)}{{\rho}c{\pi}{a_1}{a_2}}=3^22^7\Bigg\{\frac{1}{(ka_1)^{2}(ka_2)^{2}}
\\ \times \sum\limits_{m=0}^{\infty}\sum\limits_{n=0}^{\infty}\frac{\Gamma(m+n+1/2)}{m!n!\sqrt{2kd}}
\Bigg( \frac{{a_1}^m{a_2}^n}{{d}^{m+n}} \Bigg)J_{3+m}(ka_1)J_{3+n}(ka_2) \\
\times[J_{m+n+\frac{1}{2}}(kd)+i(-1)^{m+n}J_{-m-n-\frac{1}{2}}(kd)]
\Bigg\}
\end{split}
\end{equation}
where the reference is {\it rms} velocity, $\rho$ is the density
of the medium, $c$ is the speed of sound and $k$ is the
wavenumber.

%
%
%
For identical disks ($a_1=a_2=a$), an approximation for the mutual
radiation impedance is given as~\cite{oguz12}
\begin{equation}\label{eq4}
\frac{Z_{12}(ka,kd)}{{\rho}c\pi a^2}~{\approx}~A(ka)\frac{\sin(kd)
+ j\cos(kd)}{kd} \,\,\,\, \mbox{for}\,\,\,\, ka<5.5
\end{equation}
and $A(x)$ can be approximated with a $10^{th}$ order polynomial:
\begin{equation}\label{eqcited}
A(x)\approx\sum\limits_{n=0}^{10}p_{n}{x}^n
\end{equation}
The values of $p_n$'s are tabulated in Table~\ref{matProperties2}.
\begin{table}
\centering
\begin{tabular}{ccc}
  \hline
  Coefficient & Real Part & Imaginary Part \\
  \hline
  $p_{10}$ & $-1.74841\cdot 10^{-7}$ & $-2.07413\cdot 10^{-7}$\\
  $p_9$ & $-3.63059\cdot 10^{-7}$ & $~~~6.41730\cdot 10^{-6}$\\
  $p_8$ & $~~~1.00277\cdot 10^{-4}$ & $-7.76934\cdot 10^{-5}$\\
  $p_7$ & $-1.45594\cdot 10^{-3}$ & $~~~4.43450\cdot 10^{-4}$\\
  $p_6$ & $~~~8.20869\cdot 10^{-3}$ & $-1.06884\cdot 10^{-3}$ \\
  $p_5$ & $-1.52895\cdot 10^{-2}$ & $~~~5.31086\cdot 10^{-4}$ \\
  $p_4$ & $-1.28941\cdot 10^{-2}$ & $-2.40879\cdot 10^{-3}$\\
  $p_3$ & $-2.10234\cdot 10^{-2}$ & $~~~1.44157\cdot 10^{-2}$\\
  $p_2$ & $~~~2.92992\cdot 10^{-1}$ & $-7.06028\cdot 10^{-4}$\\
  $p_1$ & $-2.38707\cdot 10^{-3}$ & $~~~1.36767\cdot 10^{-4}$\\
  $p_0$ & $~~~1.73311\cdot 10^{-4}$ & $-8.68895\cdot 10^{-6}$ \\
  \hline
\end{tabular}
\caption{Coefficients of the approximate polynomial for $A(x)$}
\label{matProperties2}
\end{table}

When the disks have different radii, finding an  approximation is
harder due to the cross-coupled $ka_1$ and $ka_2$ related terms in
the mutual impedance expression. It can be numerically shown that
the mutual impedance can be written very accurately as a summation
of a separable component in $ka_1$ and $ka_2$, and an inseparable
one as follows:
\begin{equation}\label{eq5}
\frac{Z_{12}(ka_1,ka_2,kd)}{{\rho}c\pi a_1
a_2}~{\approx}~B(ka_1,ka_2)\frac{\sin(kd) + j\cos(kd)}{kd}
\end{equation}
where
\begin{equation}\label{eq55}
B(ka_1,ka_2)=S(ka_1)S(ka_2)+jI(ka_1,ka_2)
\end{equation}
where $S(x)$ is a complex-valued function with
\begin{equation}\label{eq6}
S^2(x)=A(x)
\end{equation}
With $S_r(x)$ and $S_i(x)$ real-valued functions, we write
\begin{equation}\label{eq66}
S(x)=S_r(x) + jS_i(x)
\end{equation}
From (\ref{eq6})~and~(\ref{eq66}), $S_r(x)$ and $S_i(x)$ are found
as
\begin{equation}\label{eq11}
S_r(x)=\frac{1}{\sqrt{2}}\Bigg(A_r(x)+\sqrt{A_r(x)^2+A_i(x)^2}\Bigg)^{1/2}
\end{equation}
\begin{equation}\label{eq12}
S_i(x)=\frac{A_i(x)}{\sqrt{2}(A_r(x)+\sqrt{A_r(x)^2+A_i(x)^2})^{1/2}}
\end{equation}
where $A_r(x)$ and $A_i(x)$ represent the real and imaginary parts
of $A(x)$, respectively.

The inseparable component in~(\ref{eq55}), $I(x_1,x_2)$, is
approximated as a polynomial in the following form
\begin{equation}\label{eq13}
I(x_1,x_2) \approx
(x_1-x_2)^2\sum\limits_{m=0}^{5}\sum\limits_{n=0}^{5-m}q_{mn}{x_1}^m{x_2}^n
\end{equation}
where the values of $q_{mn}$'s are given in
Table~\ref{matProperties}. We note that for $ka_1=ka_2=ka$,
(\ref{eq5}) reduces to (\ref{eq4}).

\begin{figure}[t]
\centering
\includegraphics[width=\columnwidth]{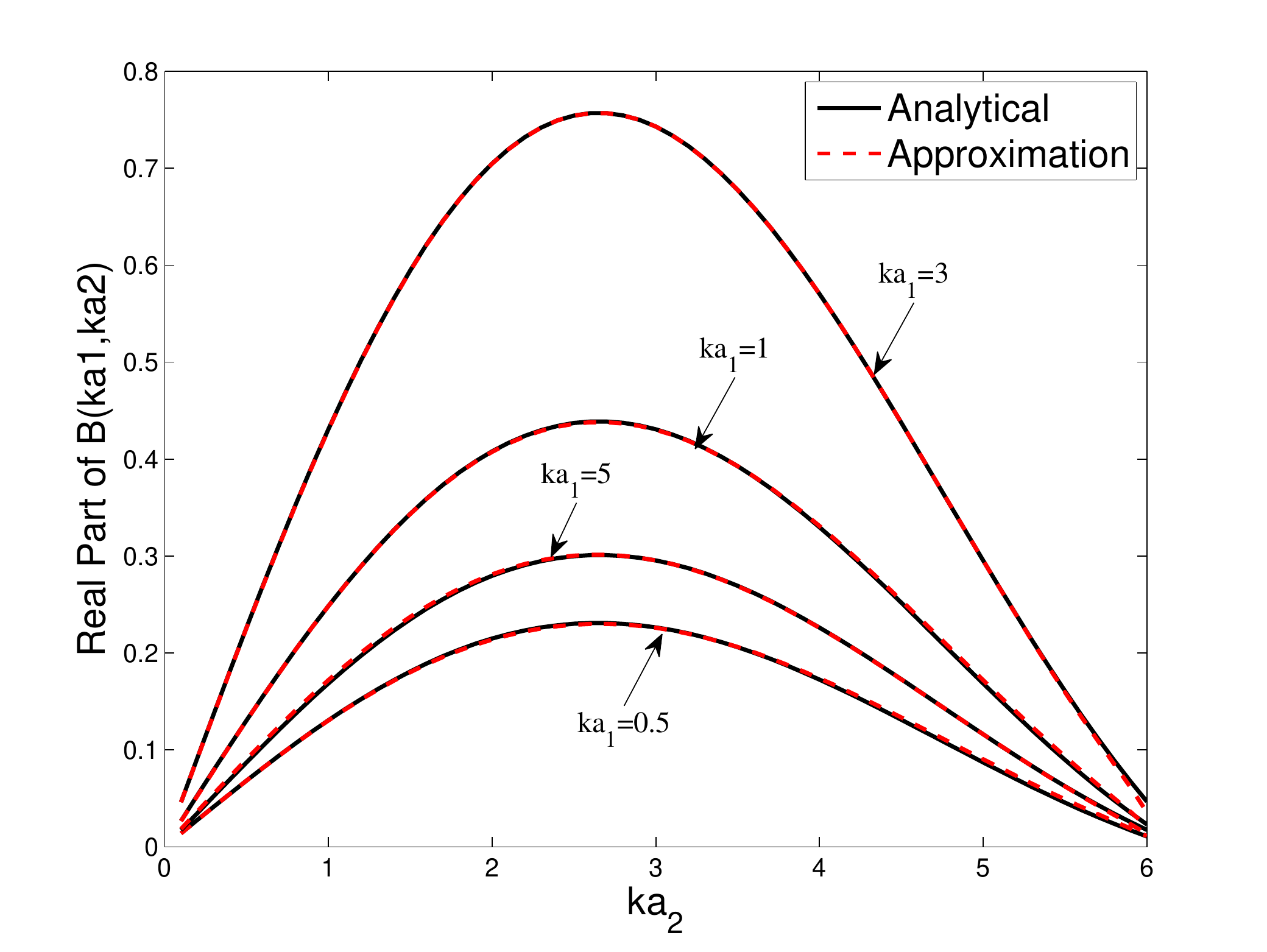}
\caption{A comparison of the real parts of the analytical  and
approximate expressions for different values of $ka_1$ and $ka_2$.
} \label{real}
\end{figure}

\begin{figure}[t]
\centering
\includegraphics[width=\columnwidth]{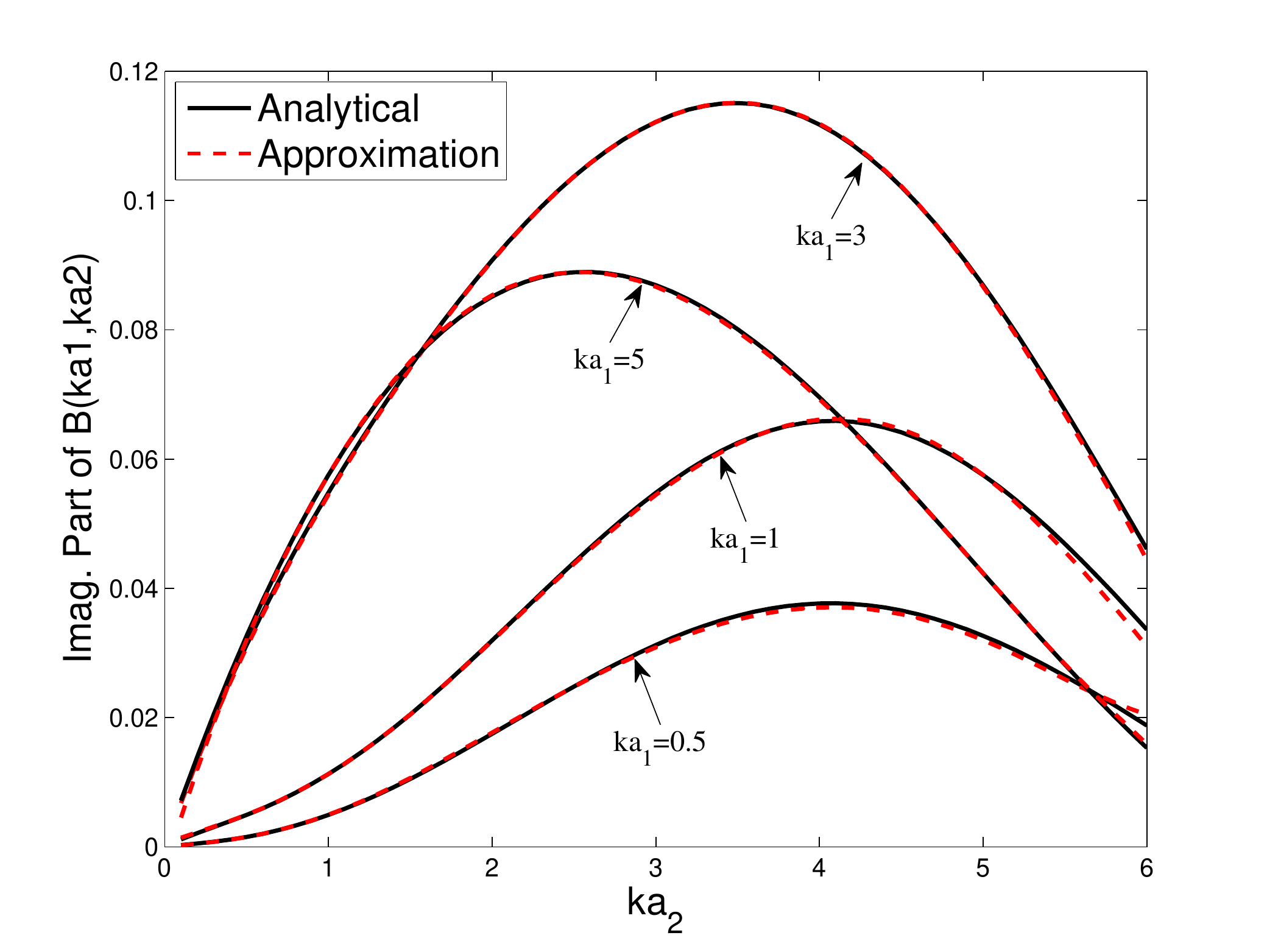}
\caption{A  comparison of the imaginary parts of the analytical
and approximate expressions for different values of $ka_1$ and
$ka_2$. } \label{imag}
\end{figure}

Figs.~\ref{real}~and~\ref{imag} are the plots of real and
imaginary parts of $B(ka_1,ka_2)$ as a function of $ka_2$ for
various values of $ka_1$. The analytical solution and the
approximate expression agree very well.
\begin{figure}[b]
\centering
\includegraphics[width=0.7\columnwidth]{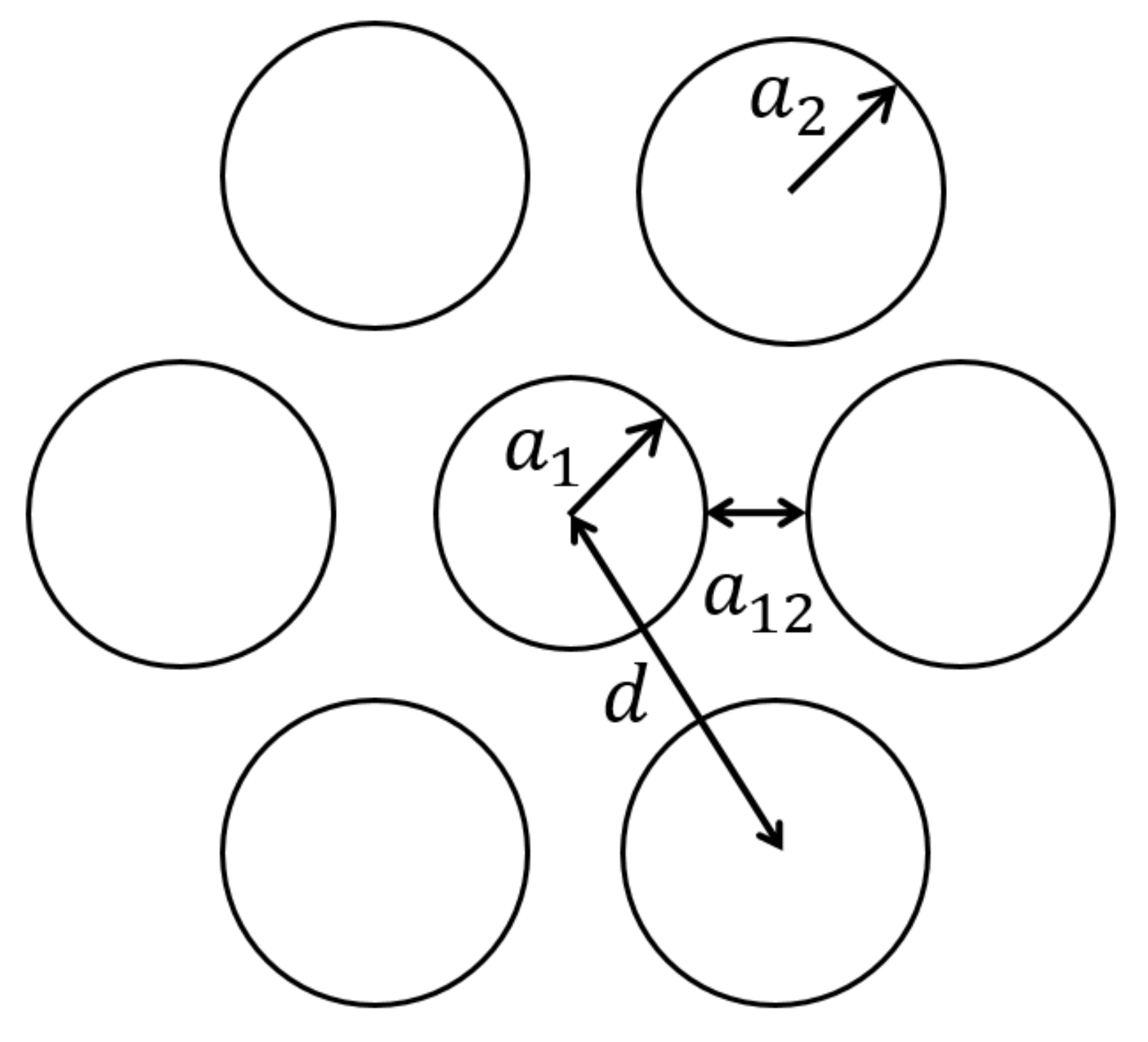}
\caption{The geometry of  CMUT element used for verification.}
\label{7Cell}
\end{figure}
\begin{table}
\centering
\begin{tabular}{lclc}
  \hline
  Coefficient & Value & Coefficient & Value \\
  \hline
  $q_{00}$ & $-1.733\cdot 10^{-3}$ & $q_{11}$ & $-8.465\cdot 10^{-5}$ \\
  $q_{01}, q_{10}$ & $~~~4.502\cdot 10^{-3}$ & $q_{12}, q_{21}$ & $~~~3.673\cdot 10^{-4}$ \\
  $q_{02}, q_{20}$ & $-3.003\cdot 10^{-3}$ & $q_{13}, q_{31}$ & $-1.475\cdot 10^{-4}$ \\
  $q_{03}, q_{30}$ & $~~~8.407\cdot 10^{-4}$ & $q_{14}, q_{41}$ & $~~~1.146\cdot 10^{-5}$ \\
  $q_{04}, q_{40}$ & $-1.091\cdot 10^{-4}$ & $q_{22}$ & $~~~1.298\cdot 10^{-5}$ \\
  $q_{05}, q_{50}$ & $~~~5.522\cdot 10^{-6}$ & $q_{23}, q_{32}$ & $~~~2.596\cdot 10^{-6}$ \\
  \hline
\end{tabular}
\caption{Coefficients of the approximate polynomial for
$I(x_1,x_2)$} \label{matProperties}
\end{table}

\section{A Comparison of the Approximation with FEM}
\begin{figure}
\centering
\includegraphics[width=\columnwidth]{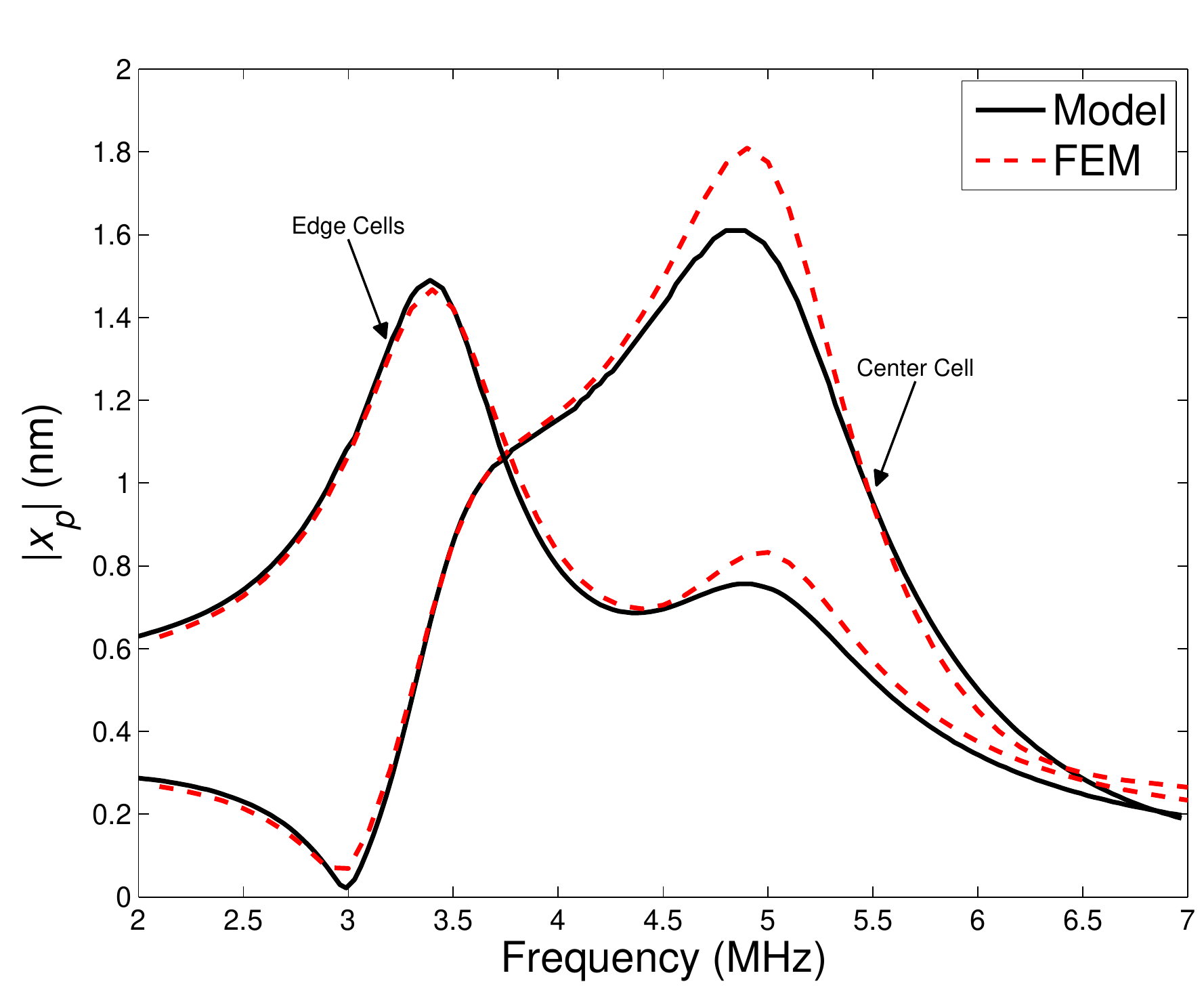}
\caption{The magnitude of the peak displacement for the center and
edge cells for an excitation voltage of 1V peak.} \label{amp}
\end{figure}

\begin{figure}
\centering
\includegraphics[width=\columnwidth]{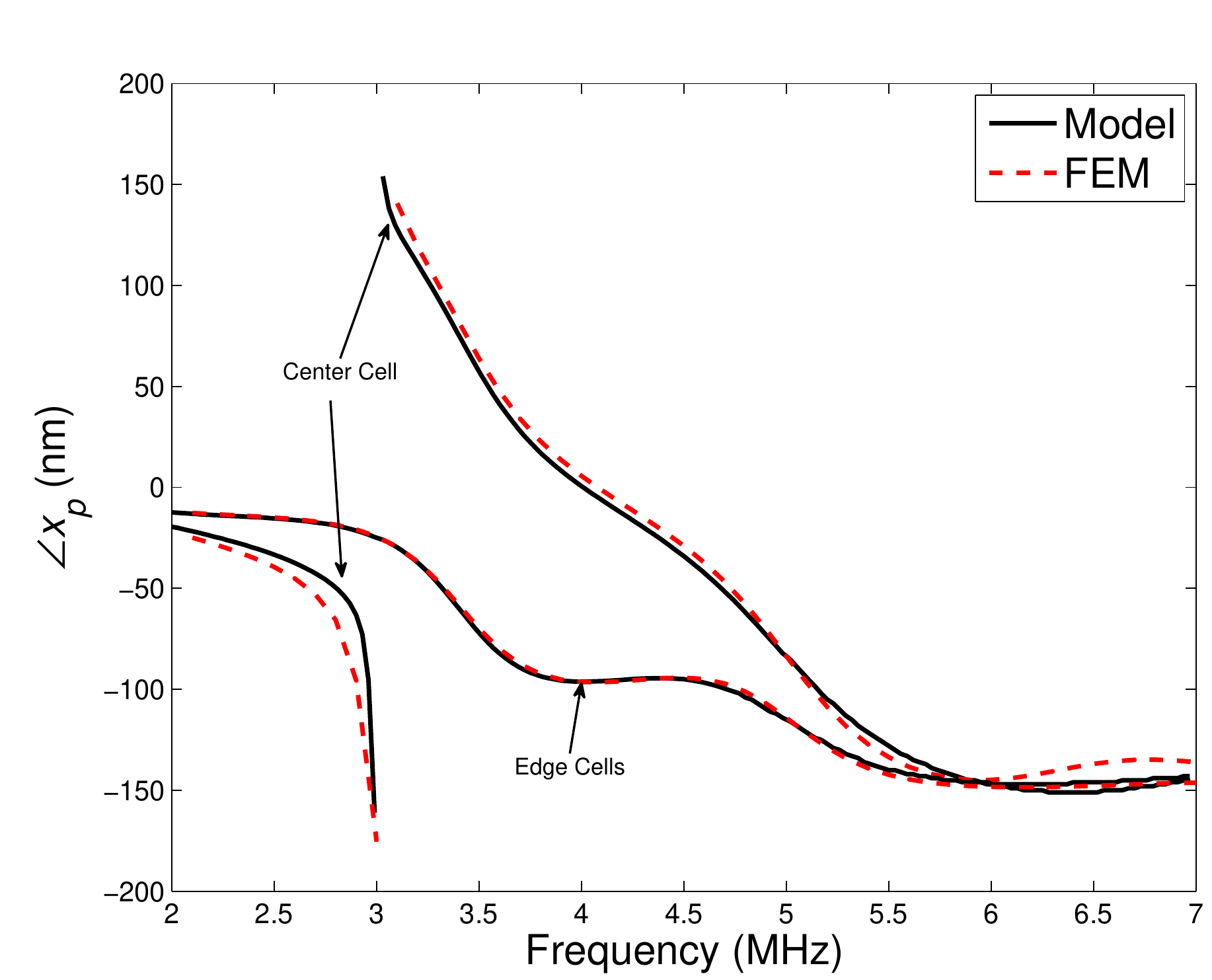}
\caption{The phase of the peak displacement for the center and edge cells.} \label{phase}
\end{figure}

The accuracy and efficiency of the presented approximation is
checked by employing it to couple CMUT cells in a cluster. We used
(\ref{eq55}) to model the mutual impedance between the central
cell and peripheral cells of the 7-cell cluster depicted
in~Fig.~\ref{7Cell} and we simulated this structure with an
electrical circuit simulator\footnote{ADS, Agilent Technologies}
capable of accepting frequency domain data.  A transient analysis
is performed using the equivalent circuit of CMUT
cell~\cite{oguz12}. Table~\ref{FEMProperties} shows the geometric
dimensions and parameters used. Equivalent circuit model
simulation results are compared with FEM analysis results in
Figs.~\ref{amp} and \ref{phase}, where a very good agreement is
observed. Notice the amplitude and phase differences of center
cell and edge cells. For example, at 2.99MHz the center CMUT cell
does not move at all, while the maximum displacements of the edge
cells and center cell occur at 3.39MHz and 4.90MHz, respectively.
At 3.75MHz, the center cell and edge cells displacement magnitudes
are equal with a 114$^\circ$  phase difference.

\begin{table}
\centering
\begin{tabular}{lc}
  \hline
  Parameter & Value \\
  \hline
  Center Cell Radius, $a_1$ & $90~{\mu}$m \\
  Edge Cell Radius, $a_2$ & $104~{\mu}$m \\
  Center-to-Edge-Cell Gap, $a_{12}$ & $24.4~{\mu}$m \\
  Gap Height, $t_{ga}$ & $150~n$m \\
  Membrane Thickness, $t_m$ & $13~{\mu}$m \\
  Insulator Thickness, $t_i$ & $100~n$m \\
  Young's Modulus, $Y_0$ & $320~$GPa \\
  Density, $\rho$ & $3270~$kg/m$^3$ \\
  Poisson Ratio, $\sigma$ & $0.263$ \\
  Bias Voltage, $V_{DC}$ & $45~$V \\
  Excitation, $V_{AC}$ & $1$V peak \\
  \hline
\end{tabular}
\caption{Parameters of the CMUT cells used in simulation}
\label{FEMProperties}
\end{table}
\section{Conclusions}
A mutual impedance approximation is presented for clamped flexural
disks having different radii. The resulting approximation is
inserted into the electrical equivalent circuit to couple CMUT
cells in a cluster and a very good agreement with the FEM results
is obtained. This approximation makes it possible to design
mixed-sized CMUT arrays using circuit simulation tools.

\bibliographystyle{ieeetran}
\bibliography{BIBalper}

\end{document}